\documentclass[aps,pre,twocolumn,showpacs,longbibliography]{revtex4-1}

\usepackage{color}

\usepackage[normalem]{ulem}
\newcommand{\stkout}[1]{\ifmmode\text{\sout{\ensuremath{#1}}}\else\sout{#1}\fi}

\usepackage{graphicx}

\begin{document}
\newcommand{\ud}{{\mathrm d}}
\newcommand{\sech}{\mathrm{sech}}

\title{Asymptotic theory of quasiperiodically driven quantum systems}

\author{David Cubero}
\email[]{dcubero@us.es}
\affiliation{Departamento de F\'{\i}sica Aplicada I, EUP, Universidad de Sevilla, Calle Virgen de \'Africa 7, 41011 Sevilla, Spain}
\author{Ferruccio Renzoni}
\email[]{f.renzoni@ucl.ac.uk}
\affiliation{Department of Physics and Astronomy, University College London, Gower Street, London WC1E 6BT, United Kingdom}


\begin{abstract}
The theoretical treatment of quasi-periodically driven quantum systems is complicated by the inapplicability of the Floquet theorem, which requires strict periodicity.
In this work we consider a quantum system driven by a bi-harmonic driving and examine its asymptotic long-time limit, the limit in which features distinguishing systems with periodic and quasi-periodic driving occur. Also, in the classical case this limit is known to exhibit universal scaling, independent of the system details,  with the system's reponse under quasi-periodic driving being described in terms of nearby periodically driven system results. We introduce a theoretical framework appropriate for the treatment of the quasi-periodically driven quantum system in the long-time limit, and derive an expression,  based on Floquet states for a periodically driven system approximating the different steps of the time evolution, for the asymptotic scaling of relevant quantities for the system at hand. These expressions are tested numerically, finding excellent agreement for the finite-time average velocity in a prototypical quantum ratchet  consisting of a space-symmetric potential and a time-asymmetric oscillating force.
\end{abstract}

\maketitle

\section{Introduction}

Quantum systems driven by a periodic perturbation are ubiquitous in physics \cite{cubren16,reimann02,hanmar09}. Their theoretical treatment is greatly simplified by the Floquet theorem, which reduces the time-dependent problem to an essentially time-independent one. Quasiperiodically driven systems are also of great interest, for the unique features they exhibit with respect to its periodic counterpart. In this case the theoretical treament is a lot more cumbersome, due to the fact that the use of Floquet theorem is strictly limited to periodic systems, and cannot be directly applied to quasiperiodically driven system. 

In this work we consider quantum systems driven by a bi-harmonic quasi-periodic drive, though the results can be easily generalized to include drives with more harmonics. We specifically examine the long-time response of the system to the driving, as it is in this limit that quasi-periodic system exhibits features distinct from their periodic counterpart. We derive an expression for the finite-time response of the system, valid in the long-time limit  and based on periodically driven states, which precisely capture all essential features of the response of the system. The theoretical framework introduced here has a broad validity, and is of direct relevance to a system consisting of a Bose-Einstein condensate driven by a bi-harmonic force, a system which has been extensively explored  theoretically \cite{denisov07a,denisov07} and experimentally \cite{salger2009,denisov13,denisov15} in the case of a periodic forcing.

 This work is organised as follows. In Sec. \ref{sec:model} we introduce the model and definitions. In Sec. \ref{sec:classical} we revisit the asymptotic theory for classical systems \cite{cubero2013,cubero2014} which will be useful as an introduction to the quantum case, fully discussed in Sec. \ref{sec:quantum}.


\section{The model}
\label{sec:model}

We consider a particle of mass $m$, subject to a spatially periodic potential $V(x)$ with period $L$, i.e. $V(x+L)=V(x)$ for all $x$.
The particle is  driven by a force depending on two frequencies, $\omega_1$ and $\omega_2$, i.e.,
\begin{equation}
F(t)=f(\omega_1 t; \omega_2t+\theta), \label{eq:force}
\end{equation}
where $ f(x_1,x_2)=f(x_1+2\pi,x_2)=f(x_1,x_2+2\pi)$ ---it is periodic in both its arguments--- and $\theta$ is the {\em driving phase}, a parameter that can be tuned to control the system's response.
We are interested in the time average of some quantity describing the particle's state, namely, the  particle's average velocity $v_{T_s}$
\begin{equation}
v_{T_s}=\frac{1}{T_s}\int_{t_0}^{t_0+T_s} dt \,v(t),
\label{eq:currentdef}
\end{equation}
where $v(t)$ is the instantaneous velocity at time $t$, $T_s$ is the observation time, and $t_0$ the initial time.  Our focus is on the asymptotic limit, in which universal scaling is expected to occur, thus the particle's velocity is computed in the infinite time limit $T_s\to\infty$. 

\section{Asymptotic theory for classical driven system}
\label{sec:classical}

In this section, we show how for the classical case, in the asymptotic limit, the finite-time average velocity of the particle under a driving with two generic --- not necessarily commensurate --- frequencies $\omega_1$ and ${\omega_2}$ can be expressed in terms of the particle's infinite-time velocity under  a {\it periodic driving}. 
In the following, we denote as $v_{pq}$ the particle's infinite-time velocity---Eq.~(\ref{eq:currentdef}) with $T_s\to\infty$---when the driving is periodic, i.e., $\omega_2=\omega_1 p/q$, with $p$ and $q$ integer coprimes.
The asymptotic theory, first derived in \cite{cubero2013}, will be then extended to the quantum regime in the following section. Precisely, we show below that in the asymptotic limit given by
\begin{equation}
T_s\to\infty, \quad (\omega_2-\omega_1p/q)T_s=\mbox{constant},
\label{eq:limit}
\end{equation}
 --- i.e. $\omega_2\to \omega_1 p/q$ --- the finite time current $v_{T_s}$ can be approximated by 
\begin{equation}
v_{T_s}\sim \frac{1}{\delta \omega_2T_s}\int_{\theta_0}^{\theta_0+\delta \omega_2 T_s}\! d\widetilde{\theta}\, v_{pq}(\widetilde{\theta}), \label{eq:asymp}
\end{equation}
where 
\begin{equation}
\delta \omega_2=\omega_2-\omega_1p/q~,
\end{equation} 
and
\begin{equation}
\theta_0=\theta+\delta\omega_2t_0~.
\end{equation} 
Equation (\ref{eq:asymp}) provides the dependence of the finite-time current on the second (or first) driving frequency, in terms of the infinite-time velocity under periodic driving.

To derive (\ref{eq:asymp}), we divide the interval between $t_0$ and $t_0+T_s$ into large sub-intervals of length $\widetilde{T}$, each containing many driving cycles, thus we can express the finite-time average velocity as 
\begin{equation}
v_{T_s}=\frac{1}{T_s}\int_{t_0}^{t_0+T_s} dt \,v(t) \sim \frac{\widetilde{T}}{T_s}\sum_j \frac{1}{\widetilde{T}}\int_{t_0+j\widetilde{T}}^{t_0+(j+1)\widetilde{T}} dt\, v(t).
\label{eq:partition}
\end{equation}
 Here $ \widetilde{T}$ is chosen as $\widetilde{T}=NT$, with $N$ a large   number, and $T=2\pi q/\omega_1$,  the driving period when $\omega_2=\omega_1 p/q$.
We can take $N$ as an integer, since in that case the deviation of $v_{T_s}$ and $v_{N\widetilde{T}}$ tends to zero for large $T_s$.
Also note that in the limit (\ref{eq:limit}) of interest here, we have $\delta\omega_2 T_s=$constant, and thus $\delta\omega_2\to0$ and $\delta \omega_2 T\to0$. In this limit we will take  $N\to\infty$, albeit with the condition  $\delta \omega_2\widetilde{T}=\delta \omega_2 N T\to 0$, which implies 
\begin{equation}
N\frac{T}{T_s}\to 0.
\end{equation}
This could be satisfied with, e.g., $N\sim (T_s/T)^{1/2}$, or $N\sim\log(T_s/T)$.
By writing $t=t_0+j\widetilde{T}+\widetilde{t}$, with $0\le\widetilde{t}\le\widetilde{T}$,  in each integral appearing in the right side of (\ref{eq:partition}), the external action driving the instantaneous  velocity $v(t)$ can be written as
\begin{eqnarray}
F(t)&=&f\left(\omega_1t;\frac{p}{q}\omega_1 t+\delta\omega_2 t+\theta\right) \nonumber \\
&=&f\left(\omega_1t;\frac{p}{q}\omega_1 t+\widetilde{\theta}_j+\delta\omega_2\widetilde{t}\right), \label{eq:driving2}
\end{eqnarray}
where $\widetilde{\theta}_j\equiv\theta_0+(\delta\omega_2 \widetilde{T}) j$.
Since $|\delta\omega_2| \widetilde{t}<|\delta\omega_2| \widetilde{T}\to0$, for small enough $|\delta\omega_2|$, the driving (\ref{eq:driving2}) in the time interval $t_j\le t\le t_j+\widetilde{T}$, with $t_j=t_0+j\widetilde{T}$, can be well approximated by 
\begin{equation}
f\left(\omega_1t;\frac{p}{q}\omega_1 t+\widetilde{\theta}_j\right),
\label{eq:approxforce}
\end{equation}
i.e., the driving associated to the periodic case $v_{pq}(\widetilde{\theta}_j)$.
In the case the velocity does not depend on the initial condition, then $v_{pq}(\theta)$ is well defined, and for large enough $T_s$ (i.e., large $\widetilde{T}$), we can approximate the finite-time integral with its value for periodic driving,
\begin{equation}
\frac{1}{\widetilde{T}}\int_{t_j}^{t_j+\widetilde{T}} dt\, v(t)\sim v_{pq}(\widetilde{\theta}_j).
\label{eq:key:approx}
\end{equation}
This expression is then exact in the limit considered here, i.e. $T_s\to\infty$, $\delta\omega_2 T_s=$constant, leading to
\begin{equation}
v_{T_s}\sim \frac{1}{\delta\omega_2T_s}\sum_j \,\Delta\widetilde{\theta} \,v_{pq}(\widetilde{\theta}_j)\sim\frac{1}{\delta\omega_2T_s}\int_{\theta_0}^{\theta_0+\delta\omega_2 T_s}\!\!d\widetilde{\theta}\,v_{pq}(\widetilde{\theta}),
\label{eq:final}
\end{equation}
where $\Delta\widetilde{\theta}=\widetilde{\theta}_j-\widetilde{\theta}_{j-1}=\delta\omega_2 \widetilde{T}$. This completes the proof for the case the  particle's velocity is independent of  the initial conditions.  Equation~(\ref{eq:asymp}) thus holds in dissipative or noisy systems where the infinite-time current $v_{pq}$ is well defined, that is, independent of the initial conditions. For a Hamiltonian classical system, additional averaging over initial conditions would be required \cite{reimann02}.


\section{Quantum systems and Floquet theorem}
\label{sec:quantum}

In non-dissipative quantum systems, there is usually a strong dependence on the initial conditions. Thus, the procedure used in the classical case does not in general apply. In particular, the approximation (\ref{eq:key:approx}), which requires the infinite-time current under periodic driving to be independent of the initial conditions, does not necessarily hold---and in consequence neither does the expression (\ref{eq:asymp}) for the finite-time current under generic driving. In this Section we show how the results for the classical case can be extended to the quantum regime. This will require taking the initial conditions into account.

\subsection{Bloch states in driven systems}
In a spatially periodic system, Bloch states form a basis of the Hilbert space. Thus we can restrict  our analysis to initial conditions consisting of a single Bloch state. The quantum dynamics for the general case can then be determined by invoking the superposition principle.

The system of interest here is the extension to the quantum case of the classical system considered previously. The quantum dynamics is described by the Hamiltonian
\begin{equation}
\mathcal{H}=\frac{p^2}{2m}+V(x)-F(t)x,
\label{eq:ham}
\end{equation}
where $p$ is the particle's momentum operator, $V(x)$ is the space-periodic potential, with period $L$,
and $F(t)$ is an arbitrary time-dependent drive, not necessarily periodic.

The Hamiltonian (\ref{eq:ham}) does not commute with the translation operator $\mathcal{T}_L$,
\begin{equation}
\mathcal{T}_L\psi(x,t)=\psi(x+L,t), \mbox{ for all $\psi(x,t)$},
\end{equation}
 because of the presence of the time-dependent drive $F(t)$.  Thus, Bloch states are not eigenfunctions of the Hamiltonian. Following a standard approach \cite{denisov07}, we perform  the following gauge transformation 
\begin{equation}
\widetilde{\psi}(x,t)=e^{-iA(t)x/\hbar} \psi(x,t), 
\label{eq:gauge}
\end{equation}
with 
\begin{equation}
A(t)=\int_{t_0}^t \!dt^\prime F(t^\prime),
\end{equation}
which yields a transformed Hamiltonian $\widetilde{\mathcal{H}}$ 
\begin{equation}
\widetilde{\mathcal{H}}=\frac{[p+A(t)]^2}{2m}+V(x),
\label{eq:ham_transformed}
\end{equation}
which commutes with $\mathcal{T}_L$.
Therefore, the system wavefunctions $\widetilde{\psi}(x,t)$ can be written as superpositions of Bloch states 
 $ \widetilde{\psi}_{k}{(x,t)}$ with quasimomentum $\hbar k$ , that is, of states with the form,
\begin{eqnarray}
\widetilde{\psi}_{k}(x,t)=e^{ik x} u_{k}(x,t), \mbox{ where } \nonumber\\
u_{k}(x+L,t)=u_{k}(x,t) \mbox{ for all $x$}.
\label{eq:psitilde}
\end{eqnarray}
The wave number $k$ can always be taken within the first Brillouin zone, i.e. $-\pi/L\le k \le \pi/L$. 


Finally,  we recall the important result that if the initial condition $\psi(x,t_0)$ is a Bloch state with wave number $k_0$, then at any time the system state $\psi(x,t)$ is also a Bloch state (i.e., an eigenstate of $\mathcal{T}_L$), but with a modified wave number given by             
\begin{equation}
k(t)=k_0+\frac{1}{\hbar}\int_{t_0}^t \!dt^{\prime}F(t^{\prime}).
\label{eq:koft}
\end{equation}
This result will be used for the calculation of the finite-time current in a quantum system. It can be readily proven by using the fact that the gauge transformed Hamiltonian $\widetilde{\mathcal{H}}$ commutes with $\mathcal{T}_L$. 

\subsection{Floquet-Bloch states}
\label{subsec:floquet}

For a time-periodic driving, indicated here by $F_{pq}$, with period $T$, i.e., $F_{pq}(t+T)=F_{pq}(t)$ for all $t$, the Floquet theorem allows one to simplify the theoretical treatment. While this does not directly apply to the case of quasi-periodic driving of interest here, Floquet states will come into play in the derivation of the velocity under quasi-periodic driving. We thus recall the main concepts which will be used in the following.

The Floquet theorem states that a periodically driven quantum system has a complete set of solutions of the form $\psi(x,t)=\exp(-i\epsilon t/\hbar)\varphi_\epsilon(x,t)$, with $\varphi_\epsilon(x,t+T)=\varphi_\epsilon(x,t)$ for all $t$, where $\epsilon$ are the quasienergies, which can be taken within the First Brillouin zone, $-\hbar\omega/2<\epsilon<\hbar\omega/2$, with $\omega=2\pi/T$.

Let us assume that the periodic driving force is unbiased,
\begin{equation}
\int_{t_0}^{t_0+T}\!\!\! dt\, F_{pq}(t)=0.
\end{equation}
Then, the gauge transformation (\ref{eq:gauge}) yields a time-periodic function $A(t)$, and thus the transformed Hamiltonian $\widetilde{\mathcal{H}}$ is also  a time-periodic ---in addition to be space-periodic. We are then entitled to write a complete set of solutions, Bloch-Floquet states,  of the form
\begin{eqnarray}
\widetilde{\psi}_{k,n}(x,t)=e^{-i\epsilon_{n}(k,\theta)t/\hbar}\widetilde{\varphi}_{k,n}(\theta,x,t), \nonumber  \\
\widetilde{\varphi}_{k,n}(\theta,x,t)=e^{ikx}\phi_{k,n}(\theta,x,t),
\label{eq:bloch-floquet}
\end{eqnarray} 
where
\begin{eqnarray}
\phi_{k,n}(\theta,x+L,t)=\phi_{k,n}(\theta,x,t) 
=\phi_{k,n}(\theta,x,t+T),\quad
\end{eqnarray} 
for all $x$ and $t$, and $n$ is an additional quantum number needed to label the states. Note that the initial condition is the same for the gauge transformed problem and the original one, $\widetilde{\psi}(x,t_0)=\psi(x,t_0)$. The functions $\phi_{k,n}(x,t)$ are space periodic, so we will assume them to be normalized within a single period, $\int_0^L dx |\phi_{k.n}(x,t)|^2=1$.

\subsection{Asymptotic finite-time velocity}

For a periodic driving, the particle velocity can be written in terms of Floquet-Bloch states, with the average current $v_n(k,\theta)$ in the $(n,k)$
Floquet-Bloch state  written as
\begin{eqnarray}
v_n(k,\theta)=\frac{1}{T}\int_{t_0}^{t_0+T}\!\!\!\! dt\int_0^L \!\!\!dx\, \psi_{k,n}(x,t)^* \nonumber \\
\times\left(-\frac{\hbar}{im}\frac{\partial}{\partial x}\right)\psi_{k,n}(x,t),
\end{eqnarray}
where we highlighted the dependence on the driving phase $\theta$.

We now consider a generic bi-harmonic driving, not necessarily periodic. We fix the frequency $\omega_1$ and consider a frequency $\omega_2$ in the neighbourhood of $\omega_1p/q$. We thus restrict the analysis to a small driving frequency deviation $\delta\omega_2=\omega_2-\omega_1p/q$, that is, within the asymptotic limit defined  by Eq. (\ref{eq:limit}).

The derivation for quasi-periodically driven quantum systems follows the same strategy as for the classical case: the interval $T_s$ is divided in many large sub-intervals of length $\widetilde{T}$, with the driving approximated by a periodic one within each sub-interval.

For the sake of simplicity, we are going to assume in the following that the initial condition $\psi(x,t_0)$ is a Bloch state with wave number $k_0$, the superposition principle allowing for generalisation. At a later time $t_j=t_0+\widetilde{T}j$, where $\widetilde{T}$ is a large multiple of $T$ (though much smaller than $T_s$), we know from (\ref{eq:koft}) that $\psi(x,t_j)$ is a Bloch state with wave number
\begin{equation} 
k_j=k_0+\int_{t_0}^{t_j}\!\! dt^\prime F(t^\prime)/\hbar.
\label{eq:koftj}
\end{equation}   
During the time interval $\widetilde{T}$ after $t_j$, the driving force $F(t)$ can be well approximated by its periodic counterpart,
\begin{equation}
F_{pq}(t)=f\left(\omega_1t;\frac{p}{q}\omega_1 t+\widetilde{\theta}_j\right),
\label{eq:Fp}
\end{equation}
corresponding to the driving phase 
\begin{equation}
\widetilde{\theta}_j=\theta+\delta\omega_2t_j.
\label{eq:thetaj}
\end{equation} 
Therefore, by using the gauge transformation 
\begin{equation}
A_j(t)=\int_{t_j}^t\!\! dt^\prime F_{pq}(t^\prime),
\label{eq:Ajt}
\end{equation}
 the state within the interval $\widetilde{T}$ can be described as a superposition of Floquet-Bloch states of the form (\ref{eq:bloch-floquet}) with $k=k_j$ and $\theta=\widetilde{\theta}_j$. Note also that since the periodic driving force is unbiased, $A_j(t)=\int_{t_0}^t\!\! dt^\prime F_{pq}(t^\prime)=A(t)$.

If $\widetilde{T}$ is large enough (see \cite{denisov07} and Appendix \ref{sec:app:coh0}), the average current within that time interval  can be approximated by their Floquet values, without interference between Floquet-Bloch states,
\begin{equation}
\frac{1}{\widetilde{T}}\int_{t_j}^{t_j+\widetilde{T}} dt\, v(t)\sim
\sum_{n}|C_n|^2 \,v_{n}\!\!\left(k_0+\int_{t_0}^{t_j}\!\! dt^\prime F(t^\prime)/\hbar,\widetilde{\theta}_j\right),
\label{eq:key:approx:quantum}
\end{equation}
where $C_n(\widetilde{\theta}_j)=\langle \widetilde{\varphi}_{k_j,n}(t_j)|\psi(t_j)\rangle$ is the projection of the system wavefunction on the $(n,k_j)$ Floquet-Bloch state of the periodic case with driving phase $\widetilde{\theta}_j$. In Appendix \ref{sec:app:coh0} we show that interference effects between Floquet-Bloch states in time-periodic systems decay as the inverse of the observation time. Furthermore,
we prove in Appendix \ref{sec:app:coh} that the absolute square of the coefficients $C_n$ remain invariant with time in the asymptotic limit considered here, that is,
\begin{equation}
|C_n(\widetilde{\theta}_j)|^2\sim |C_n(\theta_0)|^2 \mbox{ for all $j$ } (\delta\omega_2\rightarrow 0).
\label{eq:key:approx:quantum2}
\end{equation}
In other words, in this limit, the probability to be in a certain Floquet state is not affected by the probability of being in other states, remaining constant in time. 

The expressions (\ref{eq:key:approx:quantum})--(\ref{eq:key:approx:quantum2}) are the quantum case equivalent of the approximation (\ref{eq:key:approx}) derived in the classical case.
The contribution from the different sub-intervals can then be summed up, in complete analogy with the classical case, to give
\begin{eqnarray}
v_{T_s}&\sim& \frac{1}{\delta \omega_2T_s}\int_{\theta_0}^{\theta_0+\delta \omega_2 T_s}\! d\tilde{\theta}\, 
\sum_{n}|C_n|^2 \nonumber\\
& &\times  v_{n}\!\!
\left(k_0+\int_{t_0}^{t_0+T\lfloor\frac{\tilde{\theta}-\theta}{\delta\omega_2T}\rfloor}\!\! dt^\prime F(t^\prime)/\hbar,\tilde{\theta}\right),
\label{eq:final:vTs}
\end{eqnarray}
where $T=2\pi q/\omega_1$ and $|C_n|^2=|\langle\widetilde{\varphi}_{k_0,n}(t_0)|\psi(t_0)\rangle|^2$.
Therefore, in the quantum driven system, the finite-time current can be described by the current of Floquet states, with a population $|C_n|^2$ that only depends on the distribution of Floquet states at initial time. 

Equation (\ref{eq:final:vTs}) is for a system that starts with $\psi(x,t_0)$ being a Bloch state with wave number $k_0$. For an arbitrary initial condition we need only to replace $C_n$ by $C_{k_0,n}$ and add  a sum over all initial wave numbers $k_0$,
\begin{eqnarray}
v_{T_s}&\sim& \frac{1}{\delta \omega_2T_s}\int_{\theta_0}^{\theta_0+\delta \omega_2 T_s}\! d\tilde{\theta}\, 
\sum_{k_0,n}|C_{k_0,n}|^2 \nonumber\\
& &\times v_{n}\!\!
\left(k_0+\int_{t_0}^{t_0+T\lfloor\frac{\tilde{\theta}-\theta}{\delta\omega_2T}\rfloor}\!\! dt^\prime F(t^\prime)/\hbar,\tilde{\theta}\right).
\label{eq:final:vTs2}
\end{eqnarray}

\section{Numerical validation}

The validity, in the asymptotic limit, of the expressions (\ref{eq:final:vTs}) and (\ref{eq:final:vTs2}) was confirmed by  comparison with the numerical solution of the Schr\"odinger equation. 

In the numerical calculations, we have considered the following periodic potential 
\begin{equation}
V(x)=V_0\cos(2\pi x/L),
\label{eq:pot}
\end{equation}
and the following biharmonic driving force
\begin{equation}
F(t)=F_1\cos(\omega_1 t)+F_2\cos(\omega_2 t+\theta).
\label{eq:biharmF}
\end{equation}
Reduced units are defined such that $m=2\pi/L=5\hbar=1$, and initial time fixed to $t_0=0$. The time-dependent Schr\"odinger equation is solved numerically, for each initial wave number $k_0$, using a standard spectral algorithm \cite{bader11},  with a spatial mesh of 128 grid points.

In all cases, the asymptotic prediction (\ref{eq:final:vTs}) or (\ref{eq:final:vTs2}) is computed numerically by calculating the Floquet states at each $\widetilde{\theta}_j$, with a driving phase step $\Delta \widetilde{\theta}_j=0.01$. The Floquet states are determined by finding the eigenstates of the evolution operator. Figure 1 shows an example for a quantum ratchet starting from a uniform wave function, i.e. a momentum eigenstate with wave number $k_0=0$.

\begin{figure}[t]
 \includegraphics[width=6cm, clip=true]{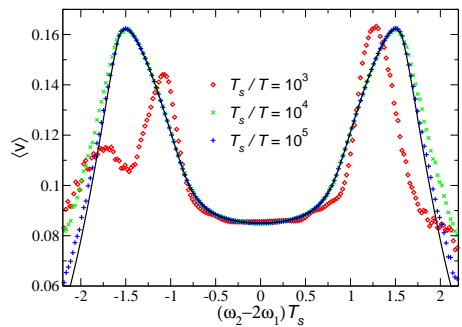}
\caption{Finite-time average velocity as a function of the second driving frequency $\omega_2$ for 
a spatially symmetric (\ref{eq:pot}) quantum ratchet subject to the time-asymmetric biharmonic force (\ref{eq:biharmF}), with $V_0=1$, $F_1=F_2=\omega_1=2$, and $\theta=-\pi/2$, 
starting from $\psi(x,0)=$constant, i.e. a momentum eigenstate with wave number $k_0=0$. 
The solid line is the prediction of the asymptotic expression (\ref{eq:final:vTs}) based on Floquet states for the periodic case $\omega_2=2\omega_1$, and the points correspond to different observation times. 
\label{fig:startfromzero}}
\end{figure}

\begin{figure}[t]
\includegraphics[width=6cm,clip=true]{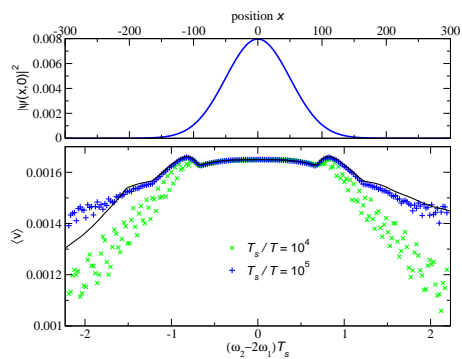}
\caption{Finite-time average velocity as a function of the second driving frequency $\omega_2$ for a system with the same parameters as in Fig.~\ref{fig:startfromzero}, starting from a Gaussian wave function, Eq.~(\ref{eq:psi0}), (the top panel shows the initial probability density in space) with average initial momentum $\hbar k_0=-0.06$.  The solid line is the prediction of the asymptotic expression (\ref{eq:final:vTs2}), and the points correspond to different  observation times.
\label{fig:super}}
\end{figure}

In Fig.~\ref{fig:super} we show the results for a quantum particle that starts from a Gaussian wave packet with a momentum distribution about $\hbar \bar{k}_0$, with $\bar{k}_0=-0.3$, and a momentum dispersion of $\sigma_p=\hbar \sigma_k$, with $\sigma_k=0.01$. More specifically, the initial condition is a superposition of $N_0=10$ momentum eigenstates between $k_{a}=-0.35$ and $k_b=-0.26$, (for reference, the minimum of the Brillouin zone is $k_{min} =-\pi/L = -0.5$),
\begin{equation}
\psi(x,0)=A_0\sum_{j=1}^{N_0}\frac{e^{i k_j x}}{\sqrt{2\pi}}\exp[-(k_j-\bar{k}_0)^2/4\sigma_k^2],
\label{eq:psi0}
\end{equation}
where $k_j=k_a+(j-1)\Delta k$, $\Delta k=(k_b-k_a)/(N_0-1)$, and $A_0$ is a normalization constant, given by $A_0=\Delta k/ (2\pi\sigma_k^2)^{1/4}$. The top panel of Fig.~\ref{fig:super} shows the spatial distribution. Despite using only 10 plane waves, the deviation from a Gaussian $|\psi(x,0)|^2=\exp[-x^2/2\sigma_x^2]/\sqrt{2\pi\sigma_x^2}$, with $\sigma_x=1/(2\sigma_k)$, is not appreciable in the plot. 
 
For both cases of an initial single Bloch state, and of a superposition state, the numerical results are in excellent agreement with the theoretical prediction, with the agreement improving for increasing interaction time, as expected. 

The results of Figs. \ref{fig:startfromzero} and \ref{fig:super} prove the validity of the asymptotic expressions (\ref{eq:final:vTs}) and (\ref{eq:final:vTs2}), thus validating our approach for the theoretical treatement of quasiperiodically driven quantum systems. 
These results also extend to the quantum regime the universal asymptotic scaling already identified in the classical case.

\section{Conclusions}
In this work we considered a quasi-periodically driven quantum system, with the driving consisting of two harmonics with incommensurate frequencies. Due to the non-periodic nature of the driving, the standard theoretical framework of Floquet states does not directly apply. We consider the asymptotic long-time limit of the system's dynamics, as it is in this limit in which quasi-periodic system exhibits features distinct from its periodic counterpart. Also, it is known from the classical case that in such a limit we can expect universal scaling, independent of the system specific details. A theoretical framework,  based on Floquet states of nearby periodic situations, which applies to quasi-periodically driven quantum systems in such a limit, is developed, and an expression for the asymptotic scaling of relevant quantities for the system at hand is derived.  These expressions were tested numerically, finding excellent agreement for the finite-time average velocity in a prototypical quantum ratchet consisting of a space-symmetric potential and a time-asymmetric oscillating force. 

\acknowledgements
We acknowledge financial support from the Ministerio de Economía y Competitividad of Spain, Grant No. FIS2016-80244-P.


\appendix 

\section{Floquet states expression of the finite-time asymptotic velocity under periodic driving}
\label{sec:app:coh0}

In this appendix we show that for a periodically driven quantum system the asymptotic velocity can be expressed as the weighted sum of the average velocity associated with the different Floquet states, each weighted with the relative population, and without interference between them. This result was first discussed in Ref.~\cite{denisov07}.  In addition, we show that the interference terms decay as the inverse of the observation time, in agreement with the observation of no interference in the infinite-time limit.

We restrict ourselves here to the periodic case, $F(t+T)=F(t)$, $T=2\pi q/\omega_1$, and consider the finite-time current
\begin{equation}
v_{T_s}=\frac{1}{T_s}\int_{t_0}^{t_0+T_s} dt \,\frac{\langle p\rangle(t)}{m},
\end{equation}
where $\langle p\rangle(t)=\langle\psi(t)|\hat{p}|\psi(t)\rangle$, $\hat{p}=-\hbar i\partial/\partial x$. The solution $|\psi(t)\rangle$ can be expanded in Floquet states $|\psi_\alpha(t)\rangle=\exp(-i\epsilon_\alpha t/\hbar)|\varphi_\alpha(t)\rangle$, where $\alpha$ is a set of quantum indexes, e.g. $\alpha=\{k,n\}$, and $|\varphi_\alpha(t+T)\rangle=|\varphi_\alpha(t)\rangle$  ---see Sec.~\ref{subsec:floquet},
\begin{equation}
|\psi(t)\rangle = \sum_{\alpha} C_\alpha |\psi_\alpha(t)\rangle.
\end{equation}
Thus
\begin{eqnarray}
v_{T_s}&=&\frac{1}{T_s}\sum_{\alpha}\sum_{\alpha^\prime}C_\alpha^* C_{\alpha^\prime}\int_{t_0}^{t_0+T_s} dt \,\nonumber\\
& &\times
\exp\left[-\frac{it(\epsilon_{\alpha^\prime}-\epsilon_\alpha)}{\hbar}\right]  \frac{p_{\alpha\alpha^\prime}(t)}{m},
\label{eq:ap:fouriertrans}
\end{eqnarray}
where $p_{\alpha\alpha^\prime}(t)=\langle\varphi_\alpha(t)|\hat{p}|\varphi_{\alpha^\prime}(t)\rangle = p_{\alpha\alpha^\prime}(t+T)$. In the limit $T_s\rightarrow\infty$, the integral in (\ref{eq:ap:fouriertrans}) becomes the Fourier transform of a periodic function, which is only different from zero at the frequencies $(\epsilon_{\alpha^\prime}-\epsilon_\alpha)/\hbar$ that are an integer multiple of $\omega=2\pi/T$. Since the quasienergies are restricted to  the first Brillouin zone,  $|\epsilon_{\alpha^\prime}-\epsilon_\alpha|<\hbar\omega$, the integral in (\ref{eq:ap:fouriertrans}) is non-zero only when $\epsilon_{\alpha^\prime}=\epsilon_{\alpha}$.

If we assume that there is no degeneracy in the distribution of Floquet states, we find
\begin{equation}
v_{T_s}\sim\sum_{\alpha}|C_\alpha^2|v_\alpha \, \mbox{ $(T_s\rightarrow\infty)$},
\label{eq:ap:vTs_final}
\end{equation}
where $v_\alpha$ is the average current associated with the Floquet state $\alpha$. This is the result used in the main body of the present work.

Note that there is usually no degeneracy for a fixed quasimomentum $\hbar k$.  Degeneracy due to states with different $k$ does not produce interference, because states with different quasimomentum are perpendicular (they are eigenstates with different eigenvalues of a Hermitian operator, a translation operator),  and the action of the momentum operator produces another Bloch state with the same quasimomentum,
\begin{eqnarray}
\hat{p}\,\psi_{,nk}(x,t)&=&\hat{p}e^{kxi}u_{k,n}(x,t)  \nonumber\\
&=&\frac{\hbar}{i} e^{kxi}\big( ki u_{k,n}(x,t)+\frac{\partial}{\partial x}u_{k,n}(x,t)\big),\,\,
\end{eqnarray}
because $u_{k,n}$ is space periodic, i.e. $u_{k,n}(x+L,t)=u_{k,n}(x,t)$, and thus so is its spatial derivative. 
Therefore, $p_{\alpha \alpha^\prime}(t)=0$ for $\alpha$ and $\alpha^\prime$ denoting Bloch states with different quasimomentum. 

Moreover, explicit analytical expressions can be easily found for the interference terms in the  periodic finite-time current (\ref{eq:ap:fouriertrans}). For two states $\alpha$ and $\alpha'$ with different quasi-energy, taking advantange of the periodicity of the functions $p_{\alpha\alpha'}(t)$, and using $T_s=N T$, we find
\begin{eqnarray}
\frac{1}{T_s}\int_{t_0}^{t_0+T_s} dt \,\exp\left[-\frac{it(\epsilon_{\alpha^\prime}-\epsilon_\alpha)}{\hbar}\right] \frac{p_{\alpha\alpha^\prime}(t)}{m}= \nonumber\\
\frac{1}{T_s}\sum_{j=0}^{N-1}\int_{t_0+T j}^{t_0+T(j+1)} dt \,\exp\left[-\frac{it(\epsilon_{\alpha^\prime}-\epsilon_\alpha)}{\hbar}\right] \frac{p_{\alpha\alpha^\prime}(t)}{m} =\nonumber \\
\frac{1}{T}\int_{t_0}^{t_0+T} dt \,\exp\left[-\frac{it(\epsilon_{\alpha^\prime}-\epsilon_\alpha)}{\hbar}\right] \frac{p_{\alpha\alpha^\prime}(t)}{m} D_{\alpha,\alpha'},\quad\quad
\label{eq:ap:cross}
\end{eqnarray}
where
\begin{equation}
D_{\alpha,\alpha'}=\frac{1}{N}\sum_{j=0}^{N-1}\exp\left[-\frac{iT(\epsilon_{\alpha^\prime}-\epsilon_\alpha)}{\hbar}\right]^j
\end{equation}
is just a geometric series, and thus easy to compute, yielding
\begin{equation}
D_{\alpha,\alpha'}=\frac{1}{N}\frac{1-\exp\left[-iTN(\epsilon_{\alpha^\prime}-\epsilon_\alpha)/\hbar\right]
}{1-\exp\left[-iT(\epsilon_{\alpha^\prime}-\epsilon_\alpha)/\hbar\right]
}.
\label{eq:ap:Dexact}
\end{equation}
Therefore, in the limit $N\to\infty$, we obtain, from (\ref{eq:ap:Dexact}), $|D_{\alpha\alpha'}|\sim 1/N$, and thus the integral (\ref{eq:ap:cross}) decays as $1/T_s$ as $T_s\rightarrow\infty$. Even if $(\epsilon_{\alpha^\prime}-\epsilon_\alpha)T/\hbar$ is very small, by expanding the exponential in the denominator, we arrive to $|D_{\alpha\alpha'}|\lesssim 2(N (\epsilon_{\alpha^\prime}-\epsilon_\alpha)T/\hbar)^{-1}$, thus the condition $(\epsilon_{\alpha^\prime}-\epsilon_\alpha)T_s/\hbar\gg 1$ guarantees  the absence of interference between those two Floquet states.
 
\section{Asymptotic time-independence of the Floquet state distribution under quasi-periodic driving}
\label{sec:app:coh}
We show here that the projection of the wave function $\psi(x,t_j)$ on the Floquet states $\widetilde{\varphi}_{k_j,n}(x,t_j)$,  i.e. $C_n(\widetilde{\theta}_j)=\langle \widetilde{\varphi}_{k_j,n}(t_j)|\psi(t_j)\rangle$, remains constant in time in absolute value, provided we are within the stated asymptotic limit, that is $\delta\omega_2\widetilde{T}\rightarrow 0$.  As we show below, this feature can be traced back to a highly oscillating behavior of $C_n$ as a function of $\widetilde{\theta}_j$. 

First, we write the evolution over a timestep using the evolution operator $U$ over the interval $t_j<t<t_{j+1}$,
\begin{equation}
|\psi(t_{j+1})\rangle=U(t_{j+1},t_j)|\psi(t_{j})\rangle.
\end{equation}
Since in this interval the driving force $F(t)$ can be well approximated by its periodic counterpart $F_{pq}(t)$, Eq.~(\ref{eq:Fp}), we can write the evolution operator in the lowest orders in $\delta\omega_2\widetilde{T}=\Delta\widetilde{\theta}$ as
\begin{equation}
U(t_{j+1},t_j)=(1+\Delta\widetilde{\theta} \,W_j)U_{pq}(t_{j+1},t_j)+\mathcal{O}(\Delta\widetilde{\theta}^2),
\label{eq:B:u}
\end{equation}
where $U_{pq}$ is the evolution operator associated with the periodic driving $F_{pq}(t)$, and $W_j$ an operator that depends on $\widetilde{\theta}_j$. It is easy to show that the fact that $U$ is an unitary operator, $U^\dag U=U^\dag U=1$, implies that $W_j$ is anti-adjoint
\begin{equation}
W_j^\dag=-W_j.
\label{eq:B:Wantiadjoint}
\end{equation}

Note that the Floquet theorem implies $|\widetilde{\varphi}_{k,n}(\theta,t_j)\rangle=|\widetilde{\varphi}_{k,n}(\theta,t_0)\rangle$. 
We recall that $|\psi(t_{j})\rangle$ is a Bloch state with a wave vector $k_j$ that depends on time,  as per Eq. (\ref{eq:koft}) or (\ref{eq:koftj}), and thus $k_j$ can be expressed as a function of the phase $\widetilde{\theta}_j$ via (\ref{eq:thetaj}).  
After the appropriate gauge transformation $A_j(t)$, as per (\ref{eq:Ajt}), we deal with Floquet-Bloch states with  $\theta = \widetilde{\theta}_j$.
We aim to express the Floquet-Bloch states $|\widetilde{\varphi}_{k_{j+1},n}(\widetilde{\theta}_{j+1},t_{0})\rangle$ at $t_{j+1}$ in terms of the states 
$|\widetilde{\varphi}_{k_{j},n}(\widetilde{\theta}_{j},t_{0})\rangle$ at time $t_j$ via a Taylor expansion in the first order in $\Delta\widetilde{\theta}$. 
Since the states  $\widetilde{\varphi}_{k(\theta),n}(\theta,x,t_0)$ are expected to vary smoothly with $\theta$, we can Taylor expand the Floquet-Bloch states at $t_{j+1}$,
\begin{eqnarray}
|\widetilde{\varphi}_{k_{j+1},n}(\widetilde{\theta}_{j+1},t_{0})\rangle=|\widetilde{\varphi}_{k_{j},n}(\widetilde{\theta}_{j},t_{0})\rangle+\nonumber\\
\Delta\widetilde{\theta} |\widetilde{\varphi}^\prime_{k_{j},n}(\widetilde{\theta}_j,t_0)\rangle +\mathcal{O}(\Delta\widetilde{\theta}^2),
\label{eq:B:varphitheta}
\end{eqnarray}
where the prime denote the derivative with respect to $\theta$ 
\begin{equation}
|\widetilde{\varphi}^\prime_{k,n}(\theta,t_0)\rangle=\frac{\partial}{\partial \theta} |\widetilde{\varphi}_{k(\theta),n}(\theta,t_0)\rangle,
\end{equation}
being in (\ref{eq:B:varphitheta}) calculated at $\theta = \widetilde{\theta}_j$. By differentiating the orthonormalization condition 
\begin{equation}
\langle\widetilde{\varphi}_{k(\theta),n}(\theta,t_{0}) | \widetilde{\varphi}_{k(\theta),n'}(\theta,t_{0})\rangle =\delta_{n,n'}, 
\end{equation}
we find
\begin{equation}
\langle\widetilde{\varphi}^\prime_{k,n}(\theta,t_{0}) | \widetilde{\varphi}_{k,n'}(\theta,t_{0})\rangle = -\langle\widetilde{\varphi}^\prime_{k,n'}(\theta,t_{0}) | \widetilde{\varphi}_{k,n}(\theta,t_{0})\rangle^*.
\label{eq:B:diffvarphi}
\end{equation}

Combining (\ref{eq:B:u}) and (\ref{eq:B:varphitheta}) into $C_n(\widetilde{\theta}_{j+1})$ yields
\begin{eqnarray}
C_n(\widetilde{\theta}_{j+1})&=&C_n(\widetilde{\theta}_{j})e^{-\frac{\epsilon_n(\widetilde{\theta}_j)\widetilde{T}i}{\hbar} }+\nonumber\\
& &\Delta\widetilde{\theta}\sum_{n'}a_{n,n'}(\widetilde{\theta}_j) C_{n'}(\widetilde{\theta}_{j})e^{-\frac{\epsilon_{n'}(\widetilde{\theta}_j)\widetilde{T}i}{\hbar} } + \mathcal{O}(\Delta\widetilde{\theta}^2), \nonumber\\
\label{eq:B:cnthetaj1}
\end{eqnarray}
where 
\begin{eqnarray}
a_{n,n'}(\widetilde{\theta}_j)&=&\langle\widetilde{\varphi}_{k_j,n}(\widetilde{\theta}_j)|W_j|\widetilde{\varphi}_{k_j,n'}(\widetilde{\theta}_j)\rangle+\nonumber\\  
& &+\langle\widetilde{\varphi}^\prime_{k,n}(\theta,t_{0}) | \widetilde{\varphi}_{k,n'}(\theta,t_{0})\rangle.
\end{eqnarray}
Equation. (\ref{eq:B:Wantiadjoint}) and (\ref{eq:B:diffvarphi}) imply
\begin{equation}
a_{n,n'}(\widetilde{\theta}_j)=-a_{n',n}(\widetilde{\theta}_j)^*.
\label{eq:B:aantiadjoint}
\end{equation}

The quantities $\epsilon_n(\widetilde{\theta}_j)\widetilde{T}/\hbar$ appearing in the exponentials of Eq.~(\ref{eq:B:cnthetaj1}) are not small, indeed they are responsible for large phase oscillations in the coefficients $C_n$ at every step. In order to obtain a smoother equation for small $\Delta\widetilde{\theta}$, we need first to transform to the following set of  coefficients $\widetilde{C}_n$, which differ from $C_n$  on a phase,
\begin{equation}
\widetilde{C}_n(\widetilde{\theta})=C_n(\widetilde{\theta})\exp{\left[i\int_{\theta_0}^{\widetilde{\theta}} \!\! d\theta'\,\,\frac{\epsilon_n(\theta')}{\hbar\delta\omega_2} \right] },
\label{eq:B:cnsmooth}
\end{equation}
yielding
\begin{eqnarray}
\widetilde{C}_n(\widetilde{\theta}_{j+1})&=&\widetilde{C}_n(\widetilde{\theta}_{j})+\Delta\widetilde{\theta}\sum_{n'}a_{n,n'}(\widetilde{\theta}_j) \widetilde{C}_{n'}(\widetilde{\theta}_{j}) \nonumber\\
& &\times e^{-\int_{\theta_0}^{\widetilde{\theta}_j+\Delta\widetilde{\theta}}\!\! d\theta' \,\,i \frac{\epsilon_{n'}(\theta')-\epsilon_{n}(\theta')}{\hbar\delta\omega_2} } + \mathcal{O}(\Delta\widetilde{\theta}^2).\quad\quad
\label{eq:B:cnthetaj2}
\end{eqnarray}
Equation (\ref{eq:B:cnthetaj2}) shows that $\widetilde{C}_n(\widetilde{\theta}_{j+1})\rightarrow \widetilde{C}_n(\widetilde{\theta}_j)$ in the limit $\Delta\widetilde{\theta}\rightarrow0$. Therefore, unlike $C_n$, the coefficients $\widetilde{C}_n$ vary smoothly at each step. After taking the limit $\Delta\widetilde{\theta}\rightarrow0$ in (\ref{eq:B:cnthetaj2}), we arrive at
\begin{equation}
\frac{\partial}{\partial\theta} \widetilde{C}_n(\widetilde{\theta}) = \sum_{n'}a_{n,n'}(\widetilde{\theta}) e^{-i\int_{\theta_0}^{\widetilde{\theta}}\!\! d\theta' \,\, \frac{\epsilon_{n'}(\theta')-\epsilon_{n}(\theta')}{\hbar\delta\omega_2} }
\widetilde{C}_{n'}(\widetilde{\theta}).
\label{eq:B:cntheta}
\end{equation}
It can be readily checked that the solution of (\ref{eq:B:cntheta}) is given by 
\begin{equation}
\widetilde{C}_n(\widetilde{\theta}) = \sum_{n'}\left( e^{S(\widetilde{\theta})}\right)_{n,n'} C_{n'}(\theta_0),
\label{eq:B:cntheta:final}
\end{equation}
where 
\begin{equation}
S(\widetilde{\theta})_{n,n'}=\int_{\theta_0}^{\widetilde{\theta}} \!\! d\theta'\,\, a_{n,n'}(\theta') 
e^{-i\int_{\theta_0}^{\theta'}\!\! d\theta'' \,\, \frac{\epsilon_{n'}(\theta'')-\epsilon_{n}(\theta'')}{\hbar\delta\omega_2}},
\label{eq:B:sdef}
\end{equation}
and we have used the fact $\widetilde{C}_n(\theta_0)=C_n(\theta_0)$. From (\ref{eq:B:aantiadjoint}), it is clear that $S$ is anti-adjoint,
\begin{equation}
S_{n,n'}(\widetilde{\theta})=-S_{n',n}(\widetilde{\theta})^*.
\label{eq:B:santiadjoint}
\end{equation}
Following the standard procedure used for Hermitian matrices, it is then easy to show that all eigenvalues of $S$ are imaginary, and its eigenvectors are, or can be chosen as, orthonormal. 

The expressions (\ref{eq:B:cntheta:final})---(\ref{eq:B:santiadjoint}) are the central results in this section. 

In the limit $\delta\omega_2\to0$, the exponential in (\ref{eq:B:sdef}) is a highly oscillatory function that makes off-diagonal terms of $S$ negligible against the diagonal terms, where the exponetial is absent. Since the diagonal terms are purely imaginary, Eq.~(\ref{eq:B:santiadjoint}), we arrive at the simple prediction
\begin{equation}
|C_n(\widetilde{\theta})|^2\sim |C_n(\theta_0)|^2 \mbox{ for all $\widetilde{\theta}$ } (\delta\omega_2\rightarrow 0).
\end{equation}
This means that the distribution of Floquet-Bloch states, i.e.,the populations $|C_n(\widetilde{\theta})|^2$, remain constant throughout the process, being unaffected by the population of different Floquet-Bloch states ---though present as a phase factor in $C_n(\widetilde{\theta})$. 
This is due to the highly oscillatory phases in the off-diagonal terms of $S$ in the limit $\delta\omega_2\rightarrow 0$.

%

\end{document}